\documentclass[12pt]{article}
\usepackage{epsfig}
\usepackage{amsfonts} 

\begin{document}

\begin{center}

{\Large THE NUMBER BEHIND THE SIMPLEST}

\

\

{\Large SIC--POVM}\footnote{Talk at the V\"axj\"o conference `Quantum and Beyond', 
June 2016.}

\vspace{18mm}

{\large Ingemar Bengtsson}\footnote{ibeng@fysik.su.se}

\vspace{15mm}

{\sl Fysikum, Stockholms Universitet,}

{\sl S-106 91 Stockholm, Sweden}

\vspace{18mm}

{\bf Abstract:}

\end{center}

\

\noindent The simple concept of a SIC poses a very deep problem in algebraic 
number theory, as soon as the dimension of Hilbert space exceeds three. A 
detailed description of the simplest possible example is given. 

\newpage

{\bf 1. Introduction to the SIC existence problem.}

\hspace{5mm}

\noindent Physicists rarely care about numbers; that is to say, they rarely 
care about the {\it nature} of numbers. Indeed, 
they routinely 
and unquestioningly rely on the real number system. Still, doubts are 
sometimes expressed. Thus, Schr\"odinger referred to quantum mechanics as a 
``makeshift'', because it seemingly does not at all challenge the notion of the 
continuum \cite{Erwin}. Be that as it may, there is an easily 
formulated quantum mechanical question forcing us to come directly to grips 
with a major unsolved problem concerning numbers.  

The question arose very naturally in the context of quantum state tomography, 
where an informationally complete and symmetric POVM would be a desirable 
thing to have \cite{Renes}, and even earlier from a highly original mathematical 
perspective on quantum theory \cite{Zauner}. It also appears in classical 
signal processing; in fact there is an engineering side of the question, 
although why this is so will not be made evident here. The question is: 
In a complex Hilbert space of $d$ 
dimensions, can one find $d^2$ vectors $|\psi_I\rangle$ such that

\begin{equation} |\langle \psi_I|\psi_J\rangle |^2 = \left\{ \begin{array}{lll}
1/(d+1) & \mbox{if} & I \neq J \\ 1 & \mbox{if} & I = J \end{array} \right. \ ? 
\label{def} \end{equation}

\noindent Such a collection of vectors, if it exists, is known as a SIC \cite{Chris}. 
SICs are also known as maximal equiangular tight frames, as minimal 
complex projective 2-designs, and---no doubt---under many other names. 

Numbers can be added and multiplied to form new numbers. Indeed they form sets, 
known as {\it fields}, that are closed under addition, subtraction, 
multiplication, and division. A standard example is the field of 
rational numbers ${\bf Q}$, which is generated from the integers by applying 
the field operations. Every field having an infinite number of elements 
contains ${\bf Q}$ as a subfield. Examples include the real number field 
${\bf R}$ (against which Schr\"odinger expressed his reservations) and the 
complex number field ${\bf C}$. The ancient Greeks (in particular, Eudoxos) 
gave a definition of 
the real numbers, but they also payed particular attention to a 
smaller extension of ${\bf Q}$, consisting of numbers that can be 
geometrically constructed using ruler and compass. Algebraically this 
corresponds to the requirement that all the numbers that occur are 
given in terms of nested square roots of rational numbers. This number 
field turned out to be too small for some purposes, for instance one 
needs the cube root $\sqrt[3]{2}$ in order to duplicate the cube. More 
generally we can consider fields built using nested {\it radicals}, including 
cube roots, quartic roots, and so on. During the Italian Renaissance there was 
a race to express the solutions of polynomial equations in terms of 
radicals, but it was eventually shown by Abel and Galois that this is 
not possible in general. The lesson learned from this brief excursion into 
history is that the choice of the number field depends on the particular task 
one is facing. 

The question is: what number field is needed to construct SICs? 
A conjectural, but precise and highly remarkable, answer is now available 
\cite{AFMY}. Our purpose here is to describe this answer using the simplest 
non-trivial example, where almost all the calculations can be done 
with pencil and paper. 

\vspace{1cm}

{\bf 2. Introducing the Weyl--Heisenberg group.}

\vspace{5mm}

\noindent If $d = 2$ the SIC existence question is trivial. Moreover, the vectors being sought 
are then always fully determined by a complete set of mutually unbiased 
bases, in a geometrically natural way. Although it is considerably harder to see 
(one will have to read several papers in order to patch a proof together 
\cite{Appleby,Szoll,Lane,Andersson}), this statement 
holds also when $d = 3$. Numerically, SICs have been found in every dimension 
where they have been looked for (this includes all dimensions $d \leq 121$, 
and a few more \cite{Scott, Andrew}), but no existence proof has been found, 
and beyond three dimensions it is very hard to see an underlying pattern in 
the solutions. Yet closer inspection 
reveals that there is a pattern also when $d \geq4$. This pattern, so far as 
it is understood, resides in number theoretical properties of exact solutions 
for SICs \cite{AFMY,AYAZ}, and the simplest non--trivial example occurs when 
$d = 4$. 

To see how exact solutions can be obtained, we first take note of the fact that 
all known SICs (with one exception) form orbits under the Weyl--Heisenberg group, 
a discrete group first brought into quantum mechanics 
by Weyl \cite{Weyl}. For $d \leq 3$ it has been proved that every SIC arises in this 
way \cite{Szoll,Lane}, and for all prime $d$ that the Weyl--Heisenberg 
group is the only possible group \cite{Zhu}. (The exception is 
generated, when $d = 8$, by another group. Like so many exceptional 
structures it is related to octonions \cite{Stacey}, and here we will assume 
that it can be left aside as a curiousity.) 

The Weyl--Heisenberg group, and the notation we use for it, need a few 
words of introduction. It is generated by two unitary operators that can 
be represented as

\begin{equation} Z|r\rangle = \omega^r|r \rangle \ , \hspace{8mm} 
X|r\rangle = |r+1\rangle \ . \end{equation}

\noindent The basis vectors are labelled by integers modulo $d$, and the phase 
factor $\omega$ is a primitive $d$th root of unity. It is convenient to introduce 
yet another phase factor $\tau = - e^{i\pi/d}$ \cite{Marcus}, and to define the 
displacement operators 

\begin{equation} D_{i,j} = \tau^{ij}X^iZ^j \ , \hspace{8mm} 0 \leq i,j < d \ 
. \end{equation}

\noindent An orbit under the Weyl--Heisenberg group is obtained by specifying a 
fiducial vector $|\psi_0\rangle$, and by forming the $d^2$ vectors 

\begin{equation} |\psi_{i,j}\rangle = D_{i,j}|\psi_0\rangle \ . \end{equation}

\noindent If we require that this orbit forms a SIC, eqs. (\ref{def}) turn into 
a set of multivariate polynomial equations for the components of the fiducial 
vector and their complex conjugates. Every solution is known numerically for 
$d \leq 50$ \cite{Scott}, and every solution is known exactly for $d \leq 21$ 
\cite{Scott,AF}. In addition, some exact solutions are known in higher dimensions. 
Beyond three dimensions the solutions are isolated, which means that the components 
are given by {\it algebraic numbers}, that is by roots of polynomials with 
integer coefficients. It is the precise nature of these numbers that is 
so surprising. 

To get these numbers out into the open, in a representation independent way, 
we proceed a little differently. A key property of the Weyl--Heisenberg group 
is that it forms a unitary operator basis, which means that any operator on 
${\bf C}^d$ can be expanded in terms of the displacement operators \cite{Schwinger}. 
In particular, 
so can the projector $|\psi_0\rangle \langle \psi_0|$. Thus  

\begin{equation} |\psi_0\rangle \langle \psi_0| = \frac{1}{d}\sum_{i,j=0}^{d-1} 
D_{i,j}^\dagger \mbox{Tr}(|\psi_0\rangle \langle \psi_0|D_{i,j}) 
= \frac{1}{d}\sum_{i,j=0}^{d-1}D_{i,j}^\dagger \langle \psi_0|D_{i,j}|\psi_0\rangle \ . 
\label{expansion} \end{equation}

\noindent It follows that the SIC fiducial can be reconstructed, uniquely up 
to an irrelevant phase, from the $d^2-1$ phase factors 

\begin{equation} e^{i\theta_{i,j}} = \sqrt{d+1}\langle \psi_0|D_{i,j}|
\psi_0\rangle \ , \hspace{8mm} (i,j) \neq (0,0) \ . \end{equation}

\noindent The number of independent phase factors is diminished by the 
symmetries possessed by the SIC, and these can be gleaned from the numerical 
solutions. 

In the simplest non--trivial example, $d = 4$, the symmetries of the SIC 
restrict the overlap phases to be 

\begin{equation} \left[ \begin{array}{cccc} \times & e^{i\theta_{0,1}} & e^{i\theta_{0,2}} 
& e^{i\theta_{0,3}} \\ e^{i\theta_{1,0}} & e^{i\theta_{1,1}} & e^{i\theta_{1,2}} 
& e^{i\theta_{1,3}} \\ e^{i\theta_{2,0}} & e^{i\theta_{2,1}} & e^{i\theta_{2,2}} 
& e^{i\theta_{2,3}} \\ e^{i\theta_{3,0}} & e^{i\theta_{3,1}} & e^{i\theta_{3,2}} 
& e^{i\theta_{3,3}} \end{array} \right]  = \left[ 
\begin{array}{rrrr} \times & u &-1 & 1/u \\ 
u & 1/u & -1/u & 1/u \\ -1 & -u & -1 & 1/u \\ 1/u & u & u & u \end{array} 
\right] \ . \label{faser} \end{equation}

\noindent Thus there is only one independent number $u$. Once this number is known, 
the entire SIC can be reconstructed from eq. (\ref{expansion}). For all $d > 4$ 
there are several independent numbers.

When $d = 4$ 
the calculations needed to compute the phase factors $e^{i\theta_{i,j}}$, starting 
from scratch by solving eqs. (\ref{def}) for a fiducial vector $|\psi_0\rangle$, 
can be done by hand. Quite easily in fact; this was described in a previous V\"axj\"o 
talk \cite{IB}. Here we simply quote the result, which is that one verifies the 
claims made so far, and moreover one finds that  

\begin{equation} u = \frac{\sqrt{5}-1}{2\sqrt{2}}+\frac{i\sqrt{\sqrt{5}+1}}{2}
\ . \label{u} \end{equation}

\noindent Now the claim is that there is the beginning of a pattern here, and 
a connection to a major unsolved problem in mathematics, that of finding numbers 
generating certain interesting number fields. This beginning, and this connection, 
are revealed once we understand the nature of the number $u$.  

\vspace{1cm}

{\bf 3. The number field of the example.}

\vspace{5mm}

\noindent Our first concern is to determine the smallest number field to which 
the number $u$ belongs. Call it ${\bf Q}(u)$, since it is an extension 
of the rational numbers. 
We will freely use the fact that the number field 
we are looking for is a subfield of the complex numbers. This gives the 
enterprise a concrete flavour, and simplifies some statements compared to 
those found in textbooks.    

Starting from $u$, we immediately conclude that $-u$, $1/u$, and $-1/u$ also 
belong to the field. So does the number 

\begin{equation} x \equiv u + 1/u = \frac{\sqrt{5}-1}{\sqrt{2}} \ . \end{equation}

\noindent In some ways $x$ is more manageable than $u$ itself, which is 
related to the fact that it is symmetric under exchanges $u \leftrightarrow 
1/u$. Going on in this way, we notice that 

\begin{equation} \sqrt{5} = 3 - \left( u + 1/u\right)^2 \hspace{12mm} 
\label{sqrt5} \end{equation}

\begin{equation} \sqrt{2} = - \frac{1}{2}\left( u + 1/u\right) \left( u - 
1/u\right)^2 \label{sqrt2} \end{equation}

\begin{equation} i\sqrt{\sqrt{5}+1} = u - 1/u \ . \hspace{10mm} \end{equation}

\noindent So these three numbers 
are in the field ${\bf Q}(u)$. Clearly $u$ can be obtained from 
them, which means that the field can be equivalently written as ${\bf Q}(u) = 
{\bf Q}(\sqrt{5}, \sqrt{2}, i\sqrt{\sqrt{5}+1})$. 
Indeed the latter three generators of the field were used in the early 
references \cite{Scott, AYAZ}. However, the field is not yet large enough to contain 
the number $i$, or the number $\tau = - (1+i)/\sqrt{2}$ that appears in the 
reconstruction formula (\ref{expansion}). We definitely want to extend our 
number field so that the number $\tau$ is included. And we want to 
do so in a principled way. This means that Galois theory must come into play. 

Introductions to Galois theory, of book length \cite{Stewart} or less than that
\cite{Hulya}, are readily available, but to really appreciate them one should look 
at a non-trivial example first---such as the one we are concerned with here. 
First we ask Mathematica for the {\it minimal polynomial}, with coefficients among the 
integers, of the algebraic number $u$. Minimal polynomials of degree $n$ always 
have $n$ distinct roots, otherwise they would not be minimal (that is, have the 
lowest possible degree). Mathematica gives the minimal polynomial for $u$ after 
only a moment's hesitation. It is of degree 8:

\begin{equation} p_1(t) = t^8 -2t^6 - 2t^4 - 2t^2 +1 \ . \end{equation}

\noindent This means that our field ${\bf Q}(u)$ can be regarded as a 
vector space over the rationals, of dimension 8, since we will perform 
our calculations modulo the equation $p_1(u) = 0$. For instance 

\begin{equation} 1/u = (1-p_1(u))/u = 2u + 2u^3 + 2u^5 - u^7 \ . \end{equation}

\noindent Every element in the field can be expressed as a polynomial in 
$u$ of a degree not exceeding seven. In general the dimension of a field, such 
as ${\bf Q}(u)$, considered as a vector space over its ground field, in this 
case ${\bf Q}$, is called its {\it degree}. So the conclusion so far is that 

\begin{equation} \{ {\bf Q}(u): {\bf Q} \} = 8 \ , \end{equation} 

\noindent where we used standard notation for the degree of an extension 
field relative to its ground field. More information about the field can 
be obtained by studying its automorphisms, 
that is to say mappings of the field onto itself which respect the field 
operations, and which leave the ground field (in this case the rationals) 
invariant. This group is known as the {\it Galois group} of the field; when 
Galois first studied it he regarded it as the group that permutes the 
roots of the minimal polynomial. The order of the Galois group equals the 
degree of the extension.  

The leading coefficient of our minimal polynomial equals 1, and it is a 
palindromic polynomial, in an obvious sense. The first property implies that $u$ 
is an {\it algebraic integer} (by definition), and the second property 
implies that $1/u$ is another root of same polynomial. It follows that both 
$u$ and $1/u$ are algebraic integers. Therefore (again by definition) $u$ is an 
algebraic {\it unit}. Another peculiarity of our polynomial is that only even 
powers appear in it, meaning that the phase factors $-u$ and $-1/u$ are roots 
of the polynomial too.  

To complete the list of phase factors appearing in eq. (\ref{faser}) we 
also need the number $-1$. It is a root of the polynomial $p_0(t) = t+1$, 
whose leading coefficient is again $1$. Therefore $-1$ is an algebraic integer 
too, and because $p_0(t)$ is palindromic it is an algebraic unit as well. In 
this context we call it a `baby unit'. Every phase factor in eq. (\ref{faser}) 
is an algebraic unit. 

Actually, the observational evidence is that the minimal polynomials of the 
SIC phases are always palindromic, for all $d$ \cite{AFMY}. This granted, 
it is easy to work out the minimal polynomial $p_1(t)$ by hand. First 
we find the minimal polynomial for the number $x = u + 1/u$, namely  

\begin{equation} p_x(t) = t^4 - 6t^2 + 4 \ . \end{equation}

\noindent Thus $x$ is an algebraic integer, but it is not a unit. 
Now we use the fact that whenever the minimal polynomial $p_{2n}$ of an 
algebraic number $z$ is a palindromic polynomial of degree $2n$, it can be obtained 
as 

\begin{equation} p_{2n}(t) = t^np_n(t+1/t) \ , \end{equation}

\noindent where $p_n(t)$ is the minimal polynomial of the number $z + 1/z$. 

But in defining the field ${\bf Q}(u)$ we seem to have stopped 
half-way: We are not able to express all the roots of the minimal polynomial. 
We have to resolve this in order to bring the full power of Galois theory 
into the play. Thus we are looking for a {\it normal} extension of ${\bf Q}$ 
allowing us to {\it split} the polynomial, that is to say we need to include 
all its roots in a field which will be larger than just ${\bf Q}(u)$. Because 
the polynomial is palindromic and depends only on even powers we know that a 
full factorization must take the form 

\begin{equation} p_1(t) = (t-u)(t+u)(t-1/u)(t+1/u)(t-r)(t+r)(t-1/r)(t+1/r) \ , 
\label{factpol4} \end{equation}

\noindent for some algebraic unit $r$. (It is a unit by construction.) Writing 
this out leads to a second degree polynomial equation for $r^2$ with coefficients 
in the field ${\bf Q}(u)$, which we can solve. But we can also simply guess the 
solution. From eq. (\ref{sqrt5}) it is evident that $\sqrt{5}$ will be left invariant 
when we permute the roots that we have so far. This suggests that it should go to 
$-\sqrt{5}$ under the exchange $u \leftrightarrow r$. If so, $r$ can be obtained 
by changing the sign in front of $\sqrt{5}$ in the expression for $u$. If we perform 
the replacements 

\begin{equation} \sqrt{5} \rightarrow - \sqrt{5} \ , \hspace{6mm} i\sqrt{\sqrt{5}+1} 
\rightarrow - \sqrt{\sqrt{5}-1} \end{equation}

\noindent we find that 

\begin{equation} r = - \frac{\sqrt{5}+1}{2\sqrt{2}} - \frac{\sqrt{\sqrt{5}-1}}{2} 
\ . \label{r} \end{equation}

\noindent We obtain $1/r$ by changing the sign in front of the second term. A direct 
calculation confirms that $r$ is a root of $p_1$. 

It is even easier to show that 

\begin{equation} \left( u + 1/u\right) \left( r + 1/r\right) = -2 \ . 
\label{ur} \end{equation}

\begin{equation} \left( u - 1/u\right) \left( r - 1/r\right) = -2i \ . 
\label{i} \end{equation}

\noindent In particular this gives the desired expression for $i = i(u,r)$. 

Now $r$ is obviously a root of 

\begin{equation} t^2 - (r+1/r)t + 1 = t^2 + \frac{2}{u+1/u}t + 1 = 0 \ . \end{equation}

\noindent Hence the minimal polynomial for $r$, with coefficients in ${\bf Q}(u)$, is 

\begin{equation} p_2(t) = t^2 + \frac{2}{u+1/u}t + 1 
= t^2 + \frac{\sqrt{5}+1}{\sqrt{2}}t + 1 \ . \end{equation}

\noindent This is a second order polynomial, which means that the degree of the 
extension from ${\bf Q}(u)$ to ${\bf Q}(u,r)$ is 2.    

The situation so far is that we have the number fields ${\bf F}_1 = {\bf Q}(u)$ 
and ${\bf F}_2 = {\bf F}_1(r) = {\bf Q}(u,r)$. The latter field is a splitting field 
of the minimal polynomial of $u$, since the polynomial admits eight roots 
over ${\bf F}_2$. The degrees of these extensions are 

\begin{equation} \{ {\bf F}_1:{\bf Q}\} = 8 \ , \hspace{5mm} \{ {\bf F}_2:{\bf F}_1\} 
= 2 \ , \hspace{5mm} \{ {\bf F}_2:{\bf Q}\} = 2\cdot 8 = 16 \ . \end{equation}

\noindent The degree is always multiplicative. As a vector space over ${\bf Q}$ the 
field ${\bf F}_2$ appears as a tensor product, with a basis consisting of the 
sixteen monomials $1, u, \dots , u^7, r, ru, \dots , ru^7$. 

The order of the Galois group equals the degree of the extension, and it permutes 
the roots of the polynomials that were used to define the field. The story becomes 
particularly simple when the extension is normal. If all the roots of the degree 
8 polynomial had been in the field obtained by adjoining one of its roots, 
the extension would have been already normal, and the Galois group would have had  
order 8. On the other hand we might have found only one root in the 
first step, and would then have been left with an irreducible polynomial of order 7. 
The second extension would then have had degree $8\cdot 7$ over the field of 
rationals. In the worst case scenario 
an extension of degree $8!$ would be needed to split the polynomial, and the Galois 
group would then be the symmetric group $S_8$. Our Galois group must have 
the comparatively modest order 16.

Three generators of our Galois group $G$ can be written down immediately. With an 
eye on the polynomial $p_2(t)$, we find them to be 

\begin{equation} g_1(u) = 1/u \hspace{1cm} g_1(r) = r \end{equation}

\begin{equation} 
g_2(u) = - u \hspace{1cm}  g_2(r) = -r \end{equation}

\begin{equation} g_3(u) = u \hspace{1cm} g_3(r) = 1/r \ . 
\end{equation}

\noindent All three generators have order 2, and together they form an abelian 
group of order 8. Let us call it $H$. We need an additional element with the property 
that $g_4(u) = r$. To see what it does to $r$ we note that 

\begin{equation} g_4(p_2) = g_4(r)^2 + \frac{2}{r+\frac{1}{r}}g_4(r) + 1 = 
g_4(r)^2 - \left( u + \frac{1}{u}\right)g_4(r) + 1 = 0 \ , \end{equation}

\noindent where eq. (\ref{ur}) was used. One solution is $g_4(r) = u$, so we choose 
the fourth generator as 

\begin{equation} \begin{array}{lll} g_4(u) = r & \ & g_4(r) = u \ . \end{array} 
\end{equation}

\noindent It is easily seen that $g_2g_4 = g_4g_2$, which means that $g_2$ belongs to 
the centre of the group. In fact $G = Z_2\times D_8$, where $D_8$ is a non-abelian 
group of order 8, easily identified with the dihedral group if we observe that 
$g_1g_4$ and $g_3g_4$ are of order 4. 

Recall that $D_8$ is the symmetry group of the square. However, we do not need to know 
this. The useful way to look at the structure of the group $G$ is to observe that the 
abelian group $H$ is a normal subgroup since 

\begin{equation} g_4g_1g_4^{-1} = g_3 \ , \hspace{7mm} g_4g_2g_4^{-1} = g_2 \ , 
\hspace{7mm} g_4g_3g_4^{-1} = g_1 \ . \end{equation}

\noindent Hence $G/H = Z_2$, which is abelian. Thus the Galois group is {\it soluble}. 
A group $G$ is called soluble if it admits a sequence of normal subgroups 
$H_k$ so that 

\begin{equation} e = H_1 \lhd H_2 \lhd H_2 \lhd \dots \lhd H_n = G \ , 
\end{equation}

\noindent where the notation is meant to imply that all the quotient groups 
$H_k/H_{k-1}$ are abelian. The name ``soluble'' is chosen because Galois 
realized that a polynomial equation can be solved 
in terms of radicals if and only if its Galois group is soluble. And indeed 
all the known SIC fiducials, although they have a complicated appearance, do share 
this remarkable feature \cite{Scott}. 

The Galois group we have arrived at admits the very short sequence 

\begin{equation} e \lhd H \lhd G \ . \end{equation}

\noindent A soluble group cannot have a shorter sequence of normal subgroups without 
actually being abelian. Similarly short sequences appear also for SICs beyond $d = 4$ 
\cite{Scott,AYAZ}.  

But we return to our example. The action of the generators of the group, on the 
numbers we have discussed, 
is worked out using eqs. (\ref{sqrt5}), (\ref{sqrt2}), and (\ref{i}). A modest amount of 
extra work is needed for the action of $g_4$. The result is given in Table 
\ref{tab:grupp}. It is interesting to observe that the generator $g_1$ effects 
complex conjugation. 

\begin{table}
\hskip 1.8cm
\begin{tabular}{|c|r|r|r|r|r|r|r|} \hline\hline 
\ & $u$ & $r$ & $\sqrt{5}$ & $\sqrt{2}$ & $i\sqrt{1+\sqrt{5}}$ & $i$ & $\tau$ \\ \hline 
$g_1$ & $1/u$ & $r$ & $\sqrt{5}$ & $\sqrt{2}$ & $-i\sqrt{1+\sqrt{5}}$ & $-i$ & $1/\tau$ \\ 
$g_2$ & $-u$ & $-r$ & $\sqrt{5}$ & $-\sqrt{2}$ & $-i\sqrt{1+\sqrt{5}}$ & $i$ & $-\tau$ \\ 
$g_3$ & $u$ & $1/r$ & $\sqrt{5}$ & $\sqrt{2}$ & $i\sqrt{1+\sqrt{5}}$ & $-i$ & $1/\tau$ \\ 
$g_4$ & $r$ & $u$ & $-\sqrt{5}$ & $-\sqrt{2}$ & $\sqrt{\sqrt{5}-1}$ & $i$ & $-\tau$ \\ \hline\hline 
\end{tabular}
\caption{{\small The generators of the Galois group, and how they act.}}
\label{tab:grupp}
\end{table}

\begin{figure}
\begin{picture}(180,115)
\newsavebox{\foldera}
\savebox{\foldera}{\hbox{\epsfig{figure=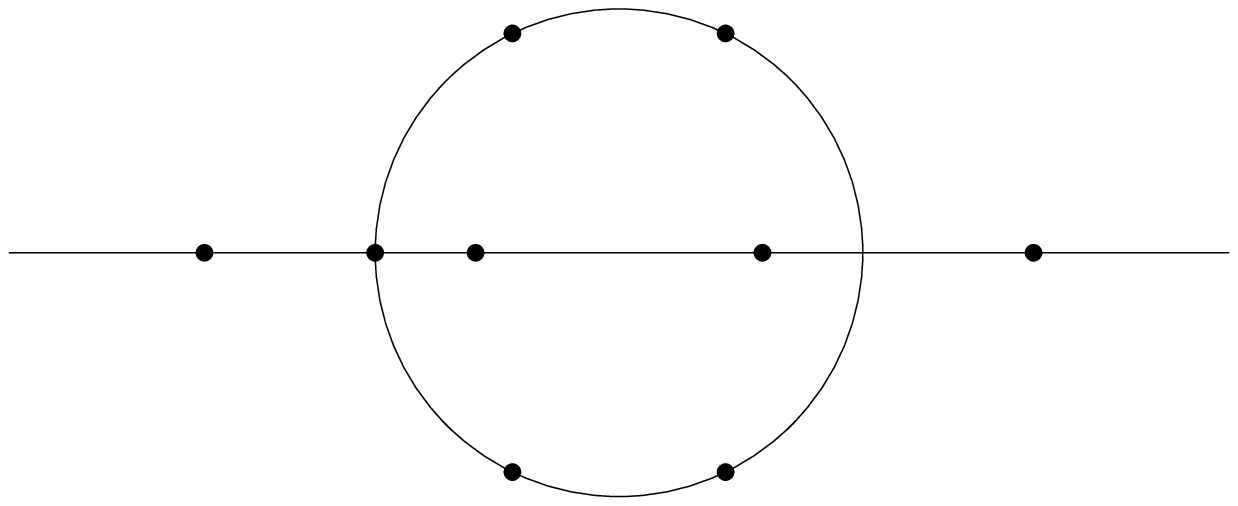,width=70mm}
}}
\put(160, 10){\usebox{\foldera}}
\put(55,1){${\bf Q}(\sqrt{5})$}
\put(47,34){${\bf Q}(\sqrt{5},\sqrt{2})$}
\put(30,75){${\bf Q}(u)$}
\put(100,75){${\bf Q}(r)$}
\put(53,111){${\bf Q}(u,r)$}
\put(72,14){\line(0,1){17}}
\put(68,47){\line(-1,1){24}}
\put(76,47){\line(1,1){24}}
\put(47,85){\line(1,1){23}}
\put(100,85){\line(-1,1){23}}
\end{picture}
\caption{{\small Left: The lattice formed by the field inclusions. Right: the algebraic 
        units we have discussed, embedded in the complex plane. On the unit circle we 
				find the phase factor $u$ and its relatives, as well as 
				the baby unit $-1$. On the real axis outside the unit circle  
				we find $r$ and its relatives.}}
\label{fig:fields}
\end{figure}

It is very interesting to observe that the subfield ${\bf Q}(\sqrt{5})$ is 
left invariant by all transformations belonging to the subgroup $H$. Hence 
the abelian group $H$ is the Galois group of the field considered as an 
extension of ${\bf Q}(\sqrt{5})$. This makes the field an {\it abelian extension} 
of the {\it real quadratic field} ${\bf Q}(\sqrt{5})$. The field ${\bf E}_{\bf R} = 
{\bf Q}(\sqrt{5}, \sqrt{2})$ is totally real, in the sense that every embedding of 
this field into ${\bf C}$ is real. This field can be extended to either 
${\bf Q}(u)$, which contains the overlap phases, or to ${\bf Q}(r)$, and the 
two of them are related by the automorphism $g_4$ when regarded as subfields of 
${\bf Q}( u,r)$, which is the field needed to construct the SIC projectors. 
Judging from the way the various patterns recur in higher dimensions 
\cite{AYAZ,AFMY}, the correct way of looking at the fields we have encountered 
is as given in Fig. \ref{fig:fields}. 

\vspace{1cm}

{\bf 4. Some further results I was told about.}

\vspace{5mm} 

\noindent We have reached the conclusion that the SIC phases are units in 
an interesting field. Now the set of units in a given field form a 
multiplicative {\it unit group}, and it is natural to ask 
how the SIC phases are positioned within that group. 

There are algorithms for computing generators of unit groups, but we have 
to take leave of pencil and paper methods at this point. A 
computer algebra package such as Magma or Sage is needed for the calculation.  
For the field ${\bf F}_1 = {\bf Q}(u)$ the unit group is 
$Z_2\times Z\times Z \times Z \times Z \times Z$. The 
finite subgroup $Z_2$ is known as the {\it torsion} subgroup, and evidently 
consists of the units $\pm 1$. A possible set of generators of the infinite 
factors is \cite{Gary} 

\begin{equation} \begin{array}{lll} u_{(1)} = 1 + \sqrt{2} & \ &  
u_{(2)} = \frac{i\sqrt{\sqrt{5}+1}}{\sqrt{2}} \\ \\
u_{(3)} = \frac{\sqrt{5}-1}{2\sqrt{2}} + \frac{i\sqrt{\sqrt{5}+1}}{2} & \ & 
u_{(4)} = \frac{\sqrt{5}-1}{\sqrt{2}} + \frac{3-\sqrt{5}}{2} 
\frac{i\sqrt{\sqrt{5}+1}}{\sqrt{2}} \\ \\ 
u_{(5)} = u_{(2)} + u_{(3)} \ . \end{array} \end{equation}

\noindent Their minimal polynomials are of degree 2, 4, 8, 8, and 8. The 
generator $u_{(1)}$ is the fundamental unit in the quadratic field 
${\bf Q}(\sqrt{2})$, while $-u_{(2)}^2$ is the fundamental unit in 
${\bf Q}(\sqrt{5})$. The generators $u_{(3)}$ and $u_{(4)}$ are complex 
phase factors. 

Remarkably, $u_{(3)}$ is equal to the SIC phase $u$. Hence the SIC phase 
has a very special position inside the
unit group. Unfortunately, for the fields arising from SICs, the unit group is known 
only in a handful of cases \cite{AFMY}. The size of 
the calculation grows quickly with the field---in fact calculating unit groups is 
one of those problems for which the best known algorithm demands a quantum 
computer \cite{Schmidt}.  

\vspace{1cm}

{\bf 5. How the example generalizes.}

\vspace{5mm}

\noindent In one sense, our example does not generalize at all: in no 
other case, except possibly for one of the two SIC fiducials in dimension 
8, can the analogous calculations be done by hand. However, this is a 
practical difficulty, not a conceptual one, and using computer algebra 
packages quite a bit of progress has been made. It is found \cite{AYAZ}, 
quite generally, that SICs give rise to fields that are abelian extensions 
of the real quadratic fields ${\bf Q}(\sqrt{D})$, where 

\begin{equation} D = (d-3)(d+1) \ . \end{equation}

\noindent (Any square factor is irrelevant, and can be divided out).  
Choosing $d > 3$ appropriately, abelian extensions of every real quadratic 
field occur in the SIC problem, and the SIC phases provide generators for their 
unit groups. A main point, emerging from recent work \cite{AFMY}, 
is that we now have a precise description of the relevant extensions. 
In all known examples they are ray class fields with conductor $d$ (or 
$2d$ if $d$ is even), or extensions thereof. These words carry deep meaning 
for algebraic number theorists. The most familiar example of a conductor is 
the integer $n$ in the phase factor $e^{2\pi i/n}$, when the rational field 
is extended to the {\it cyclotomic field} ${\bf Q}(e^{2\pi i/n})$. 
Cyclotomic fields house the most general abelian extensions of ${\bf Q}$, and 
their conductors tell us how they fit together. Considering 
abelian extensions of the imaginary quadratic fields ${\bf Q}(\sqrt{-D})$ one 
is led to replace the exponential function with special functions defined 
on suitable elliptic curves. It seems that, if we want to deal with abelian 
extensions of the real quadratic field ${\bf Q}(\sqrt{D})$, SICs provide very 
valuable insights. 
Some of the details, and how they relate to Kronecker's 
{\it Jugendtraum} and to Hilbert's unsolved 12th problem \cite{Manin}, are described 
in 
a contribution to this issue by Appleby et al. \cite{Marc}. 

Coming back to the question whether SICs exist in all dimensions, if we 
knew the field and its unit 
group, and if we had enough information about the position of the SIC phases 
within the unit group, the question might not look so formidable anymore. 

Finally, the reader may well ask for the physical significance of all this. 
The answer is not known. But the idea that quantum theory 
unquestioningly accepts the continuum has been effectively contradicted: 
elementary quantum theory seems to know some of the deepest secrets of the 
continuum.  

\newpage

\noindent \underline{Acknowledgements}: I thank the master of conference 
organizing, Andrei Khrennikov, for yet another lively week in V\"axj\"o. 
Special thanks are due to Marcus Appleby and Hulya Yadsan-Appleby for 
starting off this work as a black board session. I then relied heavily on refs. 
\cite{Andrew} and \cite{Gary}.

\end{document}